\documentclass[twocolumn]{openjournal}
\usepackage{graphicx}   
\usepackage{amsmath}    
\usepackage{amssymb,amsfonts}    
\usepackage{orcidlink}
\usepackage{verbatim}
\usepackage{hyperref}
\usepackage{xcolor}
\usepackage{longtable}
\usepackage{booktabs}

\shorttitle{Nunki is the closest core collapse progenitor candidate}
\shortauthors{Idel Waisberg \& Boaz Katz}

\begin{document}

\title{Confirming \textit{Nunki} as the closest core collapse progenitor candidate to the Sun}

\footnotetext[]{Based on observations collected at the European Southern Observatory, Chile, Program IDs 091.C-0713, 099.D-2031(A), 113.26GM.001, 114.27DS.001}

\newcommand{\weizmann}{Department of Particle Physics and Astrophysics, Weizmann Institute of Science, Rehovot 76100, Israel}

\email{email: idelwaisberg@gmail.com}

\author{\vspace{-1.2cm}Idel Waisberg\,\orcidlink{0000-0003-0304-743X}$^{1}$ \& Boaz Katz\orcidlink{0000-0003-0584-2920}$^{2}$}

\affiliation{$^1$Independent researcher, Lambda Ophiuchi Ltda}
\affiliation{$^2$\weizmann}

\begin{abstract}
We have recently suggested that \textit{Nunki}=Sigma Sagittarii is the closest core collapse progenitor candidate to the Sun based on a VLTI/GRAVITY observation that unveiled it as a $6.5+6.3 M_{\odot}$ binary at a projected separation of 0.60 au. Here we combine this observation with three VLTI/PIONIER archival and one previous MAPPIT observation to solve for the orbit of \textit{Nunki}, finding $a=1.26\pm0.05 \text{ au}$ ($P=134.779\pm0.025 \text{ days}$) and thereby confirming it as a close binary. The low orbital inclination $i=19.7\pm1.9^{\circ}$ coupled with the high projected rotational velocity $v \sin i \simeq 160 \text{ km}\text{ s}^{-1}$ and the absence of a decretion disk are a strong hint for spin-orbit misalignment. The significant eccentricity $e=0.492\pm0.003$ will cause the system to undergo eccentric Roche lobe overflow once the primary expands to $R\simeq50 R_{\odot}$, so that a merger into a $M \gtrsim 10 M_{\odot}$ star is a possible outcome. Therefore, we conclude that \textit{Nunki} at a distance $d \approx 69 \text{ pc}$ can indeed be considered the closest core collapse progenitor candidate to the Sun as it is closer than \textit{Spica} and \textit{Bellatrix} both at $d \approx 77 \text{ pc}$. Furthermore, we also report on a VLTI/GRAVITY observation of \textit{Bellatrix} that shows that it does not have any close companion with a K band flux ratio higher than 1\%; in particular, it is not a close equal mass binary as previously suspected. Two archival spectra of \textit{Nunki} illustrate how equal-mass binaries with rapidly rotating components can easily hide to become virtually spectroscopically undetectable when the radial velocity separation is several times smaller than the individual line widths.
\end{abstract}

\keywords{orbital dynamics (1184) --- stellar evolution (1599) --- stellar mergers (2157) --- optical interferometry (1168) --- stars: individual: \textit{Nunki}, \textit{Bellatrix}}

\section{Introduction}
\label{sec:introduction}

When gazing at the night sky one may wonder which is the closest star to the Sun that will eventually undergo a core collapse supernova. For a long time this title has usually been ascribed to \textit{Spica} ($\alpha$ Virginis) at a distance $d=76.7\pm4.2\text{ pc}$ \citep{vanLeeuwen2007} \footnote{We note that all the stars in this paper are very bright so that they are either unreported or unsolved in Gaia DR3.} given that its primary B1V star has a mass $M_{Aa}=11.4\pm1.2 M_{\odot}$ \citep{Tkachenko2016} that is comfortably above the threshold of about $8 M_{\odot}$ for a star to form an iron core and undergo core collapse\footnote{The fact that \textit{Spica} is a very close binary with a semi-major axis of about $30 R_{\odot}$ should cause the primary to lose most or all of its hydrogen envelope but should not prevent its ultimate core collapse.}. There are a further two stars that could possibly dethrone \textit{Spica}. $\alpha$ Arae is a B2Ve star with a mass $M \simeq 9.6 M_{\odot}$ and a rather uncertain Hipparcos distance \citep[$d=82.0_{-5.4}^{+6.1}$ pc;][]{vanLeeuwen2007}, but its parallax is likely biased by binarity and a somewhat larger distance of 105 pc has been favored from detailed modeling \citep{Meilland2007}. A more threatening claim is that of the B2III star \textit{Bellatrix} ($\gamma$ Orionis). It has a very similar distance \citep[$d=77.4 \pm 3.0$ pc;][]{vanLeeuwen2007} to \textit{Spica} but its estimated mass \citep[e.g. $8.6\pm0.3 M_{\odot}$;][]{Tetzlaff2011} is too close to the $\sim 8 M_{\odot}$ threshold for a future core collapse to be certain beyond any reasonable doubt. Its fate could also change if it were a close binary of equal components as hinted in \cite{Nieva2012} based on its spectrum potentially being ``double-lined''. Therefore, this is a good place to report that our interferometric observations of \textit{Bellatrix} show that it is a single star down to a main-sequence companion mass of at most about $1.1 M_{\odot}$ (more details can be found in Appendix \ref{app:Bellatrix}). We conclude that \textit{Bellatrix} is not a close equal mass binary and therefore its isochrone photometric mass does indeed lie just above the $\sim 8 M_{\odot}$ threshold for a future core collapse. 

In \cite{Waisberg25} we reported on the interferometric detection of a close companion to \textit{Nunki}=$\sigma$ Sagittarii\footnote{As per IAU rules the name \textit{Nunki} should formally only apply to the primary star Aa, but in this paper we will use \textit{Nunki} to refer to the binary system for simplicity.} with VLTI/GRAVITY \citep{GRAVITY17} at a projected separation of 0.60 au and with a K band flux ratio of 89\%. A close companion had been previously reported in earlier interferometric observations \citep{Hanbury74,Bedding94} but its properties could not be precisely determined. We then estimated isochrone masses of $M_{Aa} \simeq 6.5 M_{\odot}$ and $M_{Ab} \simeq 6.3 M_{\odot}$ for the primary and secondary. Based on the projected separation, we suggested that the primary should fill its Roche lobe $R_L \approx 0.38 a \sim 50 R_{\odot} \left ( \frac{a}{0.60 \text{ au}} \right )$, where $a$ is the binary semi-major axis, in the red giant branch, triggering dynamically unstable mass transfer followed by a common envelope and a merger into a $\sim 12 M_{\odot}$ star. At a distance $d=68.9\pm1.5 \text{ pc}$ that is comfortably closer than both \textit{Spica} and \textit{Bellatrix}, \textit{Nunki} could then be considered the closest progenitor candidate for a future core collapse, interestingly as the result of a merger event.

In order to put this claim into more solid footing, it is necessary to measure the orbital parameters of \textit{Nunki}. After all, there is a small albeit nonzero chance that the measured projected separation by VLTI/GRAVITY is much smaller than the true separation due to projection. A large eccentricity could also affect the evolution of the system quite substantially. 

In this paper we combine the VLTI/GRAVITY observation of \textit{Nunki} with archival VLTI/PIONIER observations (\ref{obs:pionier}) and the previously reported MAPPIT observation (\ref{mappit}) in order to measure the orbital parameters of \textit{Nunki} (\ref{orbital_parameters}). We then argue in Section \ref{sec:evolution} that the system should undergo eccentric Roche lobe overflow, which can possibly result in a common envelope event followed by a merger. We also use two archival spectra of \textit{Nunki} (\ref{obs:spectra}) to show how equal-mass binaries with fast rotating components can remain virtually undetectable in spectroscopic observations (\ref{spectral_analysis}). We briefly conclude in Section \ref{sec:conclusion}. 

\section{Data} \label{sec:data}

\subsection{VLTI/PIONIER}
\label{obs:pionier}

\textit{Nunki} was observed with the beam combiner PIONIER \citep{LeBouquin11} coupled with the 1.8-meter Auxiliary Telescopes (ATs) during three epochs in 2017 (PI: Rains). VLTI/PIONIER works in the near-infrared H band at low spectral resolution (in this case, with six spectral channels over the H band). In the first epoch, only three telescopes were used so the data consists of squared visibilities for three baselines and closure phases for one triangle. For the other two epochs, four telescopes were used leading to six baselines and four closure triangles. In all cases, the individual exposure time is 0.5 ms and the observation block consists of five files with 51,200 exposures each. In the first epoch, two blocks separated by about three hours were obtained, whereas for the other two epochs only one block was observed. A summary of the observations, including the seeing, maximum projected baseline and corresponding angular resolution, are reported in Table \ref{table:obs}. 

\begin{table*}
\centering
\caption{\label{table:obs} Summary of archival VLTI/PIONIER observations of \textit{Nunki} and best-fit binary model parameters.}
\begin{tabular}{cccccccc}
\hline \hline
\shortstack{date\\MJD} & \shortstack{seeing\\@ 500 nm (")} & \shortstack{AT config.} & \shortstack{$B_{\mathrm{proj,max}}$ (m) \\$\theta_{\mathrm{max}}$ (mas)} & calibrator(s) & \shortstack{$\frac{f_{\mathrm{A_b}}}{f_{\mathrm{A_a}}}$ (\%)\\H band} & \shortstack{$\Delta \alpha_*$\\(mas)} & \shortstack{$\Delta \delta$\\(mas)} \\ [0.3cm]
\shortstack{2017-07-16\\57950.11\\57950.23} & 0.6-1.0 & A1-G1-J3 & \shortstack{132.4\\2.4} & \shortstack{HR 6801\\HR 6766} & 0.92 & -22.76 & 11.08 \\ [0.3cm]

\shortstack{2017-08-27\\57992.06} & 0.7 & A1-G1-J2-J3 & \shortstack{132.3\\2.4} & \shortstack{HR 6801} & 0.82  & -3.58 & 19.55 \\ [0.3cm]

\shortstack{2017-09-09\\58005.07} & 1.4 & A1-G1-J2-J3 & \shortstack{132.0\\2.4} & \shortstack{HR 6766} & 0.81 & 5.05 & 14.47 \\ [0.3cm]

\hline
\end{tabular}
\end{table*}

In all three cases \textit{Nunki} was actually observed as a calibrator star to the science target $\lambda$ Sgr \citep[and it is noted as a ``failed'' calibrator due to binarity in][]{Rains20}. Fortunately, other calibrators were also observed in the same sequence, which allows us to treat \textit{Nunki} as a science target. We used the calibrator stars HR 6801 (K0III, angular diameter 1.41 mas) and/or HR 6766 (G7III, angular diameter 1.77 mas) in order to calibrate the \textit{Nunki} observations. These calibrators are located 10.2 and 10.7 deg away from \textit{Nunki} respectively and the science-calibrator observations are spaced by at most 30 min. 

We downloaded the data from the ESO archive, including the calibration files (dark and kappa matrix frames), and reduced the data using the default settings in the PIONIER data reduction software \texttt{pndrs} v3.94 \citep{LeBouquin11} downloaded from the Jean-Marie Mariotti Center (JMMC) website. The only additional step needed was to change the label of the observations of \textit{Nunki} from calibrator to science for the final calibration step. 

\subsection{Spectra}
\label{obs:spectra}

We downloaded two archival ``ready-to-use'' spectra of \textit{Nunki}. The first one is a FEROS \citep{Kaufer1999} spectrum from the ESO archive. It was taken in MJD=56524.24 with an exposure time of 25 s, wavelength coverage 3600-9000{\AA} and spectral resolution $R=48,000$. The second spectrum is from the MELCHIORS library \citep{Royer2024}, which consists of a collection of high resolution ($R=85,000$) spectra of bright stars taken with the HERMES spectrograph on the \textit{Mercator} telescope. The \textit{Nunki} spectrum was taken on MJD=55798.88 with an exposure time of 200s. The spectra were normalized by fitting a cubic spline function to carefully selected continuum regions. Both spectra are already corrected to the heliocentric frame. All wavelengths are in air. 

\section{Orbit}

\subsection{Separation vectors from PIONIER}

We fit the PIONIER interferometric data with a binary model \citep[detailed in][]{Waisberg23} consisting of the H band flux ratio ($\frac{f_{Ab}}{f_{Aa}}$) and the projected separation of Ab relative to Aa in the East and North directions $(\Delta \alpha_*, \Delta \delta)_{\mathrm{Ab,Aa}}$. The angular diameters of the stars were fixed to $\theta_{Aa} = 0.55 \text{ mas}$ and $\theta_{Ab} = 0.52 \text{ mas}$ based on the radii estimated from isochrone fitting in \cite{Waisberg25} ($R_{Aa}=4.1 R_{\odot}$ and $R_{Ab} = 3.9 R_{\odot}$) and have a negligible effect on the binary model fit since they are well below the interferometric resolution. 

\begin{figure*}[]
\centering
\includegraphics[width=\textwidth]{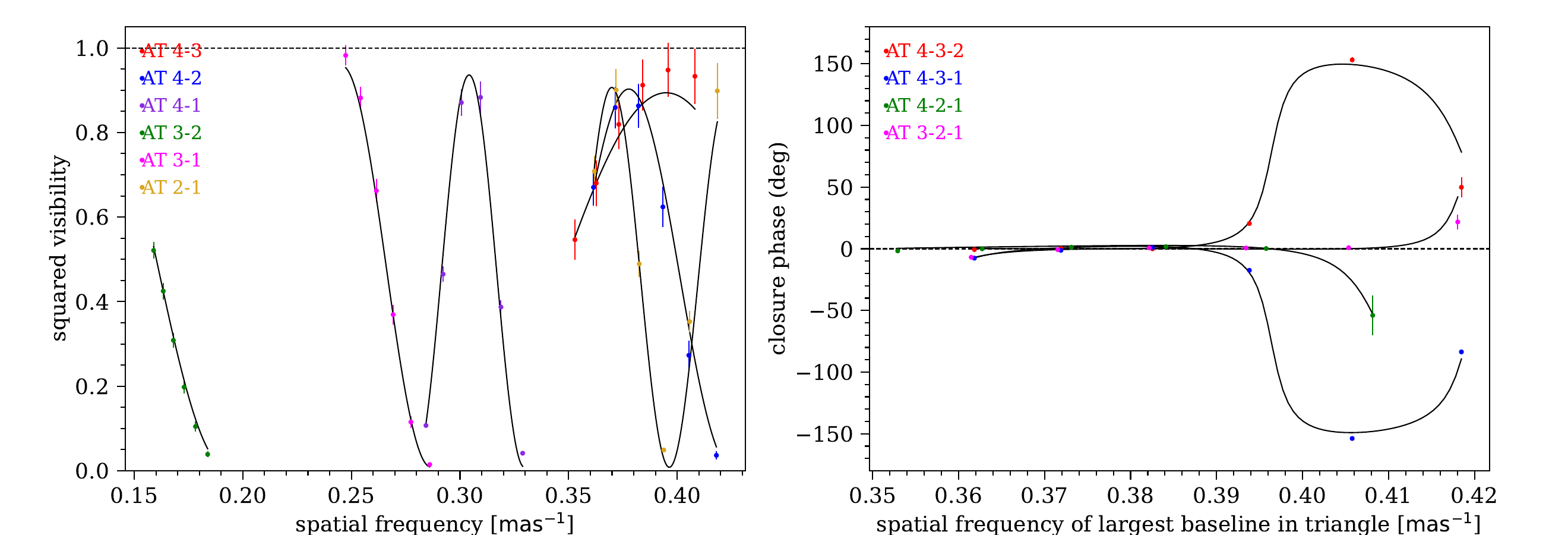}
\caption{\label{fig:pionier_fit_2017-8-27} VLTI/PIONIER data (colored) and best fit binary model (solid black) for the \textit{Nunki} data in epoch 2017-08-27. The dashed lines show the expected values for a single unresolved star.}
\end{figure*}

Figure \ref{fig:pionier_fit_2017-8-27} shows the VLTI/PIONIER data and best fit binary model for epoch 2017-08-27. Corresponding figures for the other two epochs can be found in Appendix \ref{app:pionier_fits}. The binary model fit results for each epoch are reported in Table \ref{table:obs}. We note that the H band flux ratios vary between 0.81 and 0.92 (suggesting an error of about 5\% per epoch) but fixing them to the K band ratio from the VLTI/GRAVITY observation (89\%) has a negligible effect on the separation compared to the error. The formal astrometric errors from interferometric binary fits are usually underestimated due systematic errors from calibration and correlations between spectral channels. We therefore adopt a conservative error of 0.1 mas for the PIONIER positions as in \cite{LeBouquin13}. 

\subsection{Additional separation vectors}
\label{mappit}

In addition to the three VLTI/PIONIER points from 2017, we also make use of the VLTI/GRAVITY measurement taken in 2024 \citep{Waisberg25}, for which we adopt an error of 0.020 mas in each direction based on the astrometric error derived from orbital fitting of VLTI/GRAVITY data on bright binaries \citep[e.g.][]{Waisberg25b}. Furthermore, we also adopt the position vector measured with MAPPIT in 1991 (MJD $\simeq$48463.5) reported in \cite{Bedding94}. Their reported separation $\rho=11.5\pm2 \text{ mas}$ and position angle $202^{\degr} \pm 10^{\degr}$ translate to a separation vector $(\Delta \alpha_*, \Delta \delta) = (-4.3 \pm 2.0, -10.7 \pm 2.0) \text{ mas}$. We note that their measurement assumed a magnitude difference $\Delta V = 0$ which is close to $\Delta K = 0.13$ measured by VLTI/GRAVITY (the stars are hot enough that $\Delta V \simeq \Delta K$). Although much more uncertain than the modern measurements, this data point still plays an important role in constraining the orbit given the low number of measurements available. 

\subsection{Orbital parameters}
\label{orbital_parameters}

\begin{figure}[]
\centering
\includegraphics[width=0.5\textwidth]{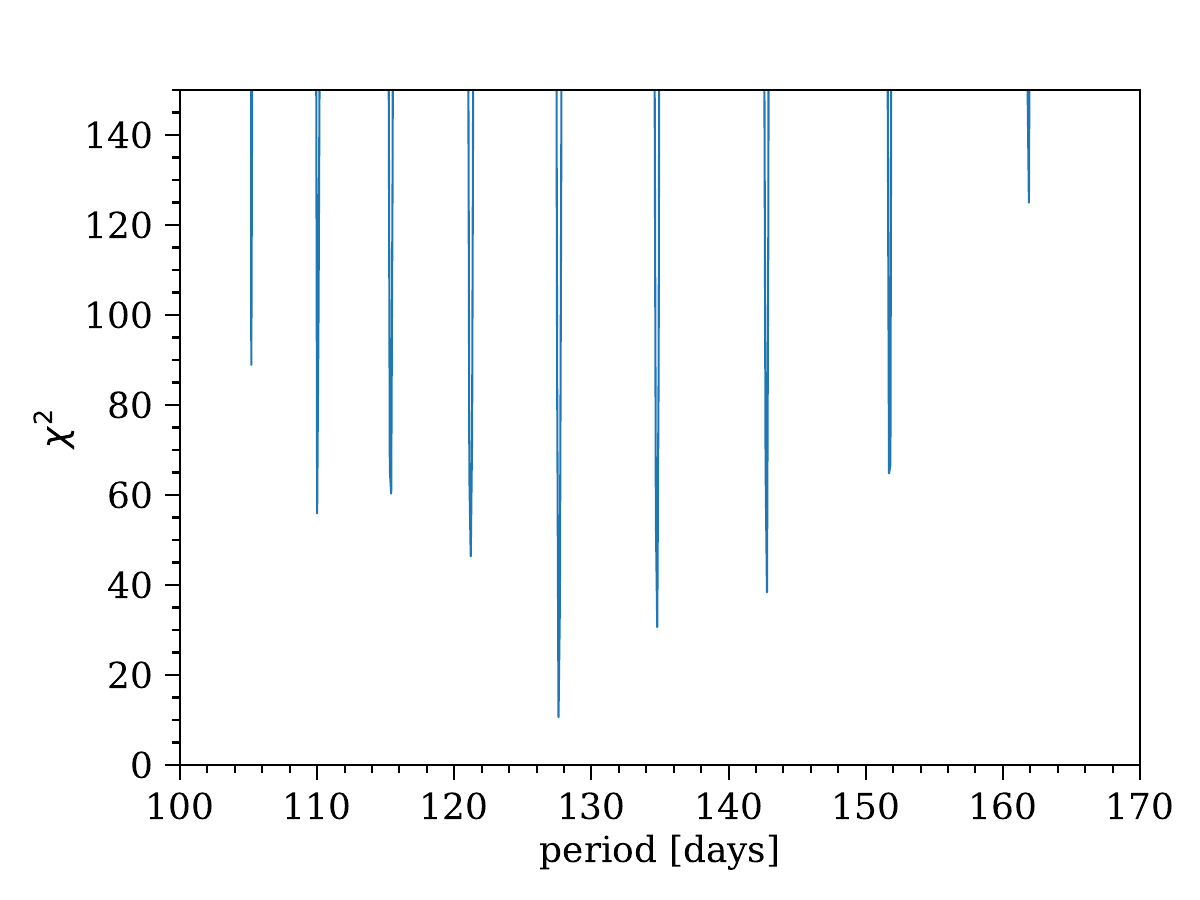}
\caption{\label{fig:period_minima} Minimum $\chi^2$ (sum of squared weighted residuals) over the ($P$,$e$,$T_p$) grid as a function of orbital period.}
\end{figure}

To find the orbital parameters of \textit{Nunki}, we created a three-dimensional grid in orbital period ($P$) between 100 and 200 days with steps of 0.1 day, eccentricity ($e$) between 0 and 1 with steps of 0.02 and time of periastron ($T_p$) in steps of 0.1 day. At each point in the grid, the optimal Thiele-Innes elements $A,B,F,G$ (which map to the remaining orbital parameters, namely semi-major axis $a$, orbital inclination $i$, argument of pericenter $\omega$ and longitude of the ascending node $\Omega$) can be quickly computed by linear regression according to

\begin{align}
\Delta \alpha_* = B X + G Y \\
\Delta \delta = A X + F Y \\
X = \cos E - e \\ 
Y = (1-e^2)^{1/2} \sin E 
\end{align}

\noindent where $E$ is the eccentric anomaly computed from Kepler's Equation. Figure \ref{fig:period_minima} shows the minimum $\chi^2$ (sum of squared weighted residuals) over the grid as a function of period. There are several local minima due to the large time gap between the different data sets. 

For each of the local minima, we then ran a non-linear least squares fit with all orbital parameters free to vary with a starting position set by the local minima. We found a clear global minimum for an orbital period $P=134.8 \text{ d}$ ($\chi^2 = 0.23$) \footnote{Note that this is not the global minimum in Figure \ref{fig:period_minima}, which illustrates the need to explore each local minima.}. The two adjacent minima at $P=127.6 \text{ d}$ and $P=142.8 \text{ d}$ have substantially worse $\chi^2$ (5.7 and 10.5 respectively) and can be rather safely excluded. Note that the low $\chi^2$ of the best fit solution (corresponding to a $\chi^2_{\mathrm{red}} = 0.077$) does not necessarily imply that the astrometric errorbars are significantly underestimated because given the low number of degrees of freedom (10 measurements - 7 parameters = 3) the resulting $\chi^2$ distribution has a mean of 3 and a large standard deviation of $\sqrt{2 \times 3} = 2.45$. 

\begin{figure*}[]
\centering
\includegraphics[width=\textwidth]{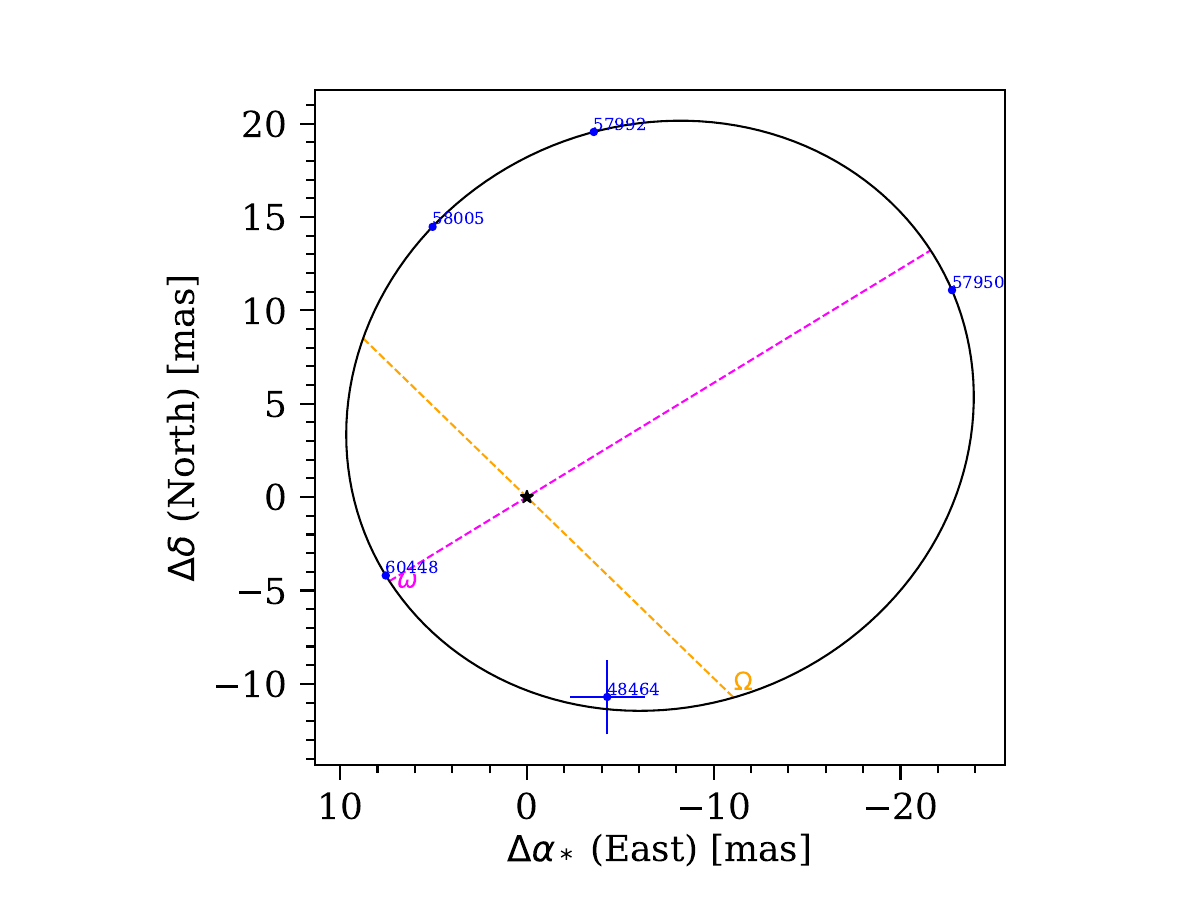}
\caption{\label{fig:orbital_fit} Astrometric data (blue) labeled by MJD and best fit orbit (black) for \textit{Nunki} = Sigma Sagittarii. The dashed magenta and orange lines show the line of apsides and the line of nodes respectively.}
\end{figure*}

Figure \ref{fig:orbital_fit} shows the measured astrometric positions (also labeled by their MJD) together with the best fit orbit. The best fit parameters and their uncertainties are reported in Table \ref{table:orbital_fits}. The uncertainties (defined as the 2.3\% and 97.7\% percentiles) were estimated by generating 20,000 resamples of the data according to the errors and refitting the orbit to find the distributions of the orbital parameters, which are shown in Figure \ref{fig:corner} in Appendix \ref{app:corner_plot}. We note that the lack of radial velocities lead to a perfect degeneracy between $(\Omega, \omega)$ and $(\Omega + 180^{\degr}, \omega + 180^{\degr})$, so we report the solution with $\Omega < 180^{\degr}$ as is standard practice. 

We find that the orbit is rather eccentric with $e=0.492\pm0.003$ and a semi-major axis $a=18.0 \pm 0.2 \text{ mas} \leftrightarrow 1.25 \pm 0.05 \text{ au}$, corresponding to a pericenter distance $a_p = 0.64 \pm 0.03 \text{ au} \leftrightarrow 138 \pm 7 R_{\odot}$. The VLTI/GRAVITY observation reported in \cite{Waisberg25} coincidentally happened very close to the pericenter passage and so the projected separation underestimated the true semi-major axis by a factor of two despite the orbit being close to face on. 

\begin{table}
\centering
\caption{\label{table:orbital_fits} Best fit Keplerian parameters for \textit{Nunki}.}
\begin{tabular}{cc}
\hline \hline

\shortstack{a\\(mas)\\(au)} & \shortstack{$18.0\pm0.2$\\$1.26\pm0.05$} \\ [0.3cm]

e & $0.492\pm0.003$ \\ [0.3cm]

\shortstack{$i$\\(deg)} & $19.7\pm1.9$ \\ [0.3cm]

\shortstack{$\Omega$\\(deg)} & $45.9\pm8.0$ \\ [0.3cm]

\shortstack{$\omega$\\(deg)} & $76.4\pm7.1$ \\ [0.3cm]

\shortstack{P\\(days)} & $134.779\pm0.025$ \\ [0.3cm]

\shortstack{$T_p$\\(MJD)} & $48588.1 \pm 2.2$ \\ [0.3cm]

\hline \\ [0.1cm]

\shortstack{$a_p$\\(au)} & $0.64\pm0.03$ \\ [0.3cm]

\shortstack{$M_{\mathrm{dyn}}$\\($M_{\odot}$)} & $14.6 \pm 1.8$ \\ [0.3cm]

\hline
\end{tabular}
\tablenotetext{0}{Notes:}
\tablenotetext{0}{The uncertainties correspond to the 2.3\% and 97.7\% percentiles of the parameter distributions.}
\tablenotetext{0}{For the physical semi-major axis and dynamical mass, the adopted distance is $d=68.9\pm3.0 \text{ pc}$ ($2 \sigma$).}
\tablenotetext{0}{Alternative solution with $(\Omega, \omega) \leftrightarrow (\Omega + 180^{\degr} , \omega + 180^{\degr})$ is possible.}
\end{table}

\subsection{Spectral analysis}
\label{spectral_analysis}

\begin{figure*}[]
\centering
\includegraphics[width=\textwidth]{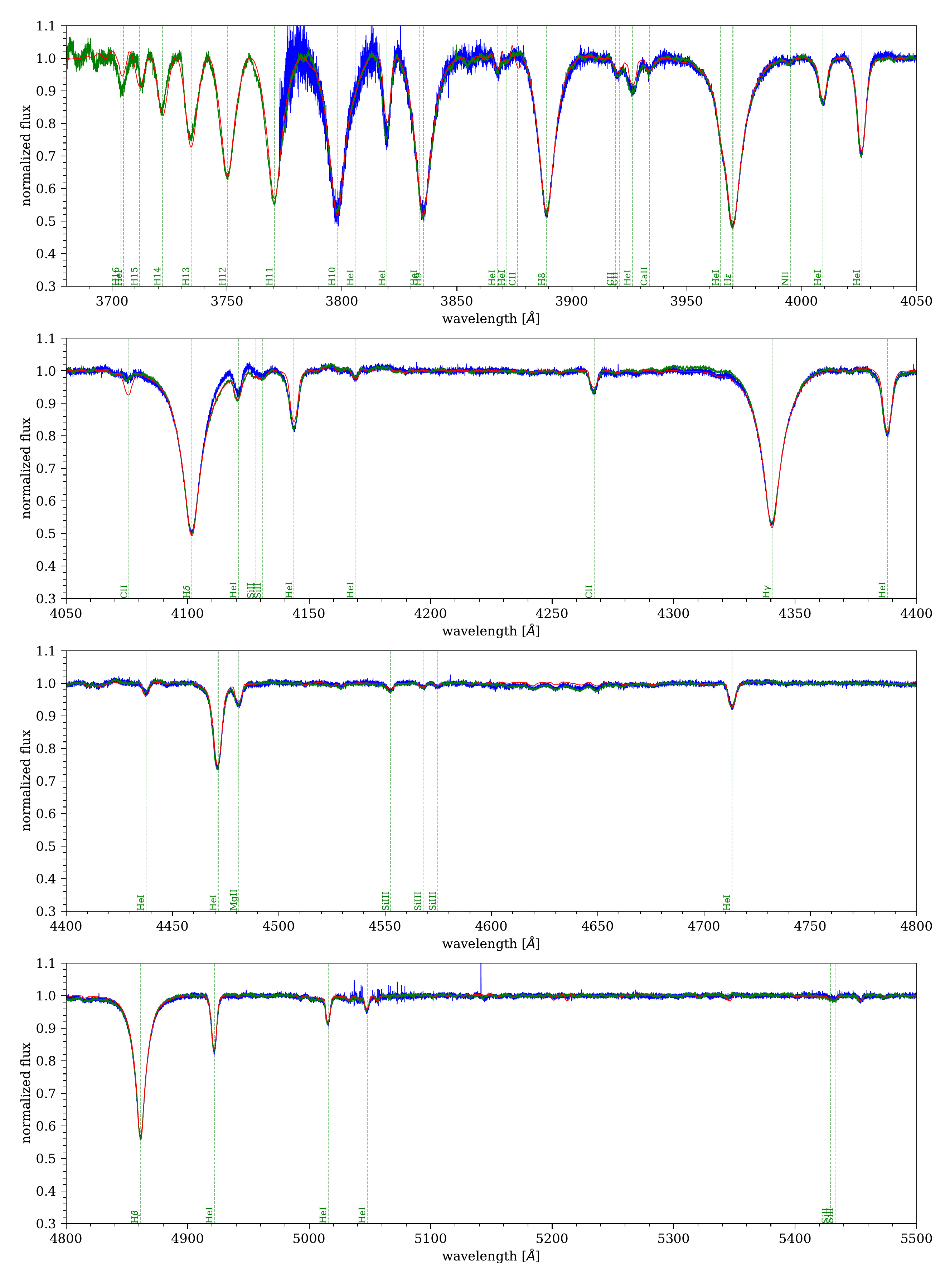}\\
\caption{\label{fig:plot_spectrum_1} Spectral atlas of \textit{Nunki}. The FEROS spectrum is shown in green and the HERMES spectrum in blue. The red line shows a BSTAR2006 model with $T_{\mathrm{eff}}=19 \text{ kK}$, $\log g=4.0$ and $v \sin i = 160 \text{ km}\text{ s}^{-1}$.}
\end{figure*}

\begin{figure*}[]
\centering
\includegraphics[width=\textwidth]{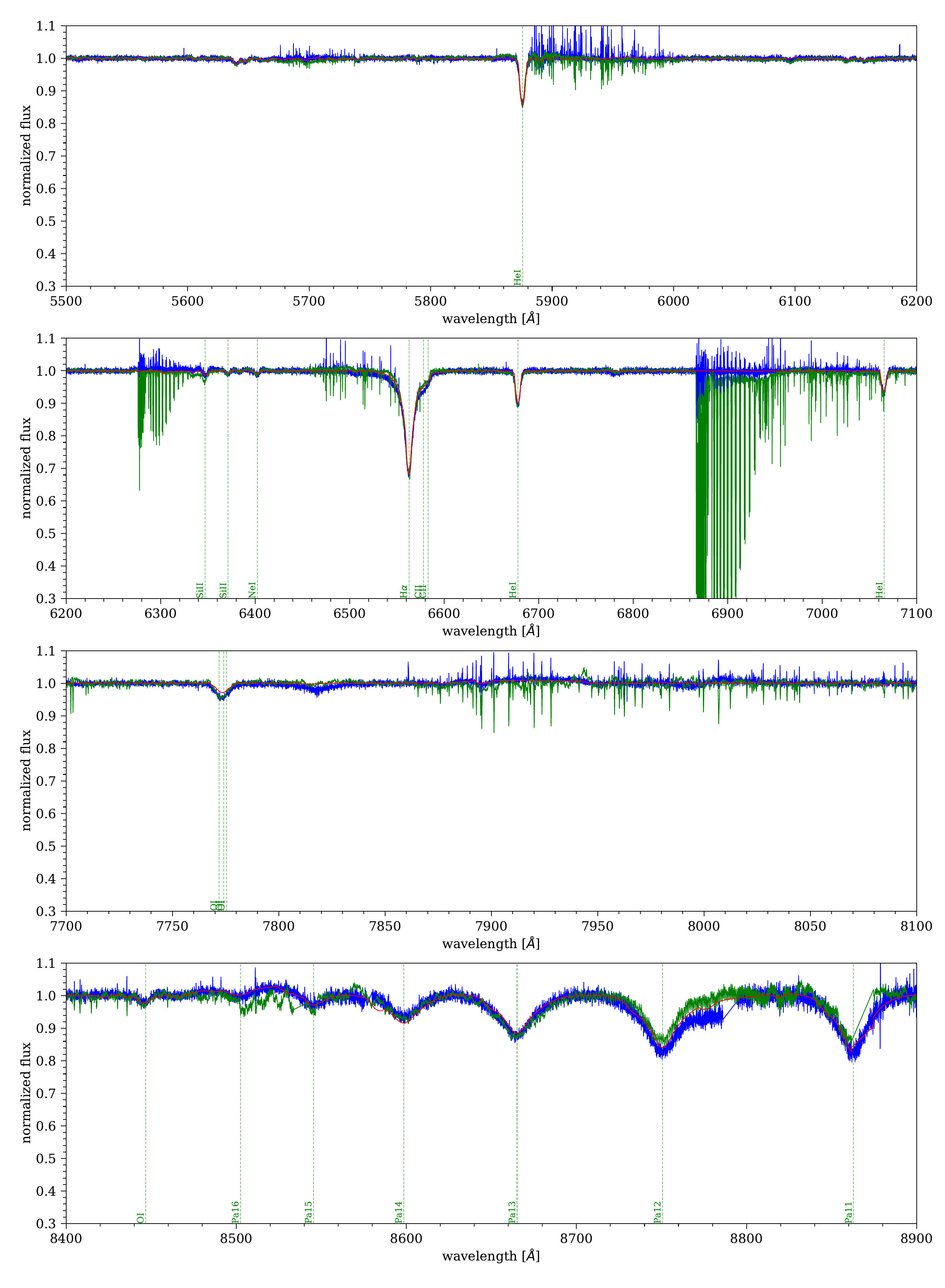}\\
\caption{\label{fig:plot_spectrum_2} Continuation of Figure \ref{fig:plot_spectrum_1}. Starting at about 5900{\AA} there are strong telluric features.}
\end{figure*}

Figures \ref{fig:plot_spectrum_1} and \ref{fig:plot_spectrum_2} show the FEROS (green) and HERMES (blue) spectra of \textit{Nunki}. They agree very well and there is no sign of variability. The dominant lines are those of HI and HeI, but there many weak metal lines due to CII, NII, OI, NeI, MgII, SiII, SiIII and CaII. The most prominent lines are marked and are listed in Table \ref{table:line_list}. 

There is no evidence for double lines in the spectrum but this is hardly surprising because the lines are very broad, with metal lines (for which pressure broadening is very weak) having a Full Width Half Maximum of 230 $\text{ km}\text{ s}^{-1}$, which is several times larger than the expected radial velocity separation between the two components. In fact, based on the orbital solution determined in \ref{orbital_parameters} the radial velocity difference of the two components has a semi-amplitude 

\begin{align}
| K_{Aa} - K_{Ab} | = \frac{2 \pi a}{\sqrt{1-e^2}} \sin i \simeq 40 \text{ km}\text{ s}^{-1}
\end{align}

Specifically, for the FEROS and HERMES epochs the orbital solution predicts radial velocity separations $| v_{z,Aa} - v_{z,Ab} | \simeq 41 \text{ km}\text{ s}^{-1}$ and $5 \text{ km}\text{ s}^{-1}$, respectively. The fact that the two spectra are indistinguishable (even their subtraction does not reveal signs of binarity) is a nice illustration of how easy it is to spectroscopically hide an equal-mass binary with fast rotating components when their radial velocity separation is several times smaller than the individual line widths. 

We estimated the effective temperature by comparing the Equivalent Width ratio between SiII (4128.05{\AA}, 4130.88{\AA}) and SiIII (4552.62{\AA}, 4567.82{\AA}, 4574.76{\AA}) lines with those in those in the publicly available NLTE, line-blanketed model atmosphere grid BSTAR2006 \citep{Lanz2007}. The grid covers the temperature range 15-30 kK in steps of 1kK and $\log g$ 1.75-4.75 in steps of 0.25 and has a solar abundance and a microturbulent velocity $\xi_t = 2 \text{ km} \text{ s}^{-1}$. We use $\log g=4.0$ appropriate for the masses and radii for the stars in \text{Nunki} and found $T_{\mathrm{eff}} \simeq 18.5 \text{kK}$, in very good agreement with previous spectrophotometric estimates \citep{Zorec2009} and the B2.5V spectral classification \citep{Morgan1979}. 

Previous estimates of the projected rotational velocity of \textit{Nunki} found $v \sin i \sim 160 \text{ km}\text{ s}^{-1}$ \citep{Slettebak1975,Abt2002}. The red line in Figures \ref{fig:plot_spectrum_1} and \ref{fig:plot_spectrum_2} shows a BSTAR2006 model with $T_{\mathrm{eff}}=19 \text{ kK}$, $\log g=4.0$ and $v \sin i = 160 \text{ km}\text{ s}^{-1}$, confirming that the latter provides a good match to the line shapes.

\section{Discussion}

\subsection{Dynamical mass}

The dynamical mass $M_{\mathrm{dyn}} = 14.6 \pm 1.8 M_{\odot}$ (2$\sigma$ error) from the orbital solution is consistent within $2\sigma$ with the total isochrone mass $M_{\mathrm{iso}} \simeq 12.8 M_{\odot}$ estimated in \cite{Waisberg25}. The Hipparcos parallax is $p = 14.32 \pm 0.29 \text{ mas} \leftrightarrow d=69.8 \pm 1.5 \text{ pc}$ \citep[1$\sigma$ errors;][]{vanLeeuwen2007}; a true parallax of 14.9 mas (d=66.8 pc) would hardly affect the isochrone mass and would bring the dynamical mass in agreement with it. We note that given that the orbital period is a significant fraction of one year, it is possible that the orbital motion could slightly bias the parallax on the order of the photocenter semi-major axis 

\begin{align}
a_{\mathrm{phot}} = a_{Aa} \frac{1-f/q}{1+f} = \frac{q}{1+q} a \frac{1-f/q}{1+f} \simeq 0.4 \text{ mas}
\end{align}

\noindent where $f=0.877$ is the flux ratio and $q=0.97$ is the mass ratio. 

Given that the isochrone masses are also a better match for the spectroscopic temperature $T\simeq18.5 \text{ kK}$ (main-sequence $7.3 M_{\odot}$ stars would be expected to be somewhat hotter with $T\simeq20.5 \text{ kK}$), in the following we will adopt the isochrone masses $M_{Aa} \simeq 6.5 M_{\odot}$ and $M_{Ab} \simeq 6.3 M_{\odot}$ as the more reasonable estimates. 

\subsection{A hint for spin-orbit misalignment}

The low orbital inclination $i=19.7\pm1.9^{\degr}$ is interesting considering the relatively high projected rotational velocity of \textit{Nunki} $v \sin i \simeq 160 \text{ km}\text{ s}^{-1}$. If the stellar spin axes were aligned with the orbital axis, this would imply a very high rotational velocity $v \simeq 490 \text{ km}\text{ s}^{-1}$, which is about $90\%$ of the critical velocity $\left ( \frac{GM}{R} \right )^{1/2}$ for $M=6.5 M_{\odot}$ and $R=4.0 R_{\odot}$. If this were the case one might expect that \textit{Nunki} would be a Be star. Since there is no evidence of a face-on decretion disk (such as emission lines and near-infrared excess), this is probably an interesting clue for a significant spin-orbit misalignment. In fact, given the large eccentricity and the fact that the stars are rotating much faster 

\begin{align}
P_{rot} \lesssim \frac{2 \pi R}{v \sin i} \simeq 1.3 \text{ day}
\end{align}

\noindent than the pseudo-synchronization orbital period, there is no strong reason to expect that the stellar spin axes would be aligned with the orbital axis. 

\subsection{Future evolution} \label{sec:evolution}

Armed with precise orbital parameters, we are now on more solid ground to assess the future evolution of \textit{Nunki} and specifically whether it can be considered a core collapse progenitor candidate. 

There will be strong interaction through mass transfer once the primary star expands to a radius comparable to the equivalent Roche lobe radius at the periastron separation $r_p = a (1-e) = 0.64 \text{ au} \leftrightarrow 138 R_{\odot}$:

\begin{align}
R_{Aa} \simeq f(q) a = 0.38 a = 52 R_{\odot} \\
f(q) = \frac{0.49 q^{-2/3}}{0.6 q^{-2/3} + \ln (1 + q^{-1/3})}
\end{align}

\noindent \citep{Eggleton1983}. Using the MESA Isochrones and Stellar Tracks \citep[\texttt{MIST};][]{Choi16,Dotter16,Paxton11,Paxton13,Paxton15} Equivalent Evolutionary Points (EEP) file for a $6.6 M_{\odot}$ star, we find that it will reach such a radii in 25 Myr with an effective temperature $T_{Aa} \simeq 5,700 \text{ K}$, luminosity $L_{Aa}=2600 L_{\odot}$ and a He core mass of $0.9 M_{\odot}$. Such an ``early-RGB'' star should still have an envelope that is almost entirely radiative with only a very shallow convective outer layer. Meanwhile, at this point the secondary (simulated as a $6.4 M_{\odot}$ at the corresponding age using the corresponding \texttt{MIST} evolution file) has just slightly expanded to $R_{Ab} \simeq 7 R_{\odot}$ and has a luminosity $L_{Ab} \simeq 2600 L_{\odot}$ similar to the primary. 

If the orbit were circular, there are two stabilizing factors that would favor thermal-timescale stable mass transfer, namely a mass ratio close to unity \citep[e.g.][]{Soberman1997} and a donor star envelope that is mostly radiative and therefore tends to contract adiabatically upon mass loss \citep[e.g.][]{Hjellming1987, Pavlovskii2015}. In this case, as the mass ratio inverts the orbit would actually be expected to expand and even detach upon conservative mass transfer. 

However, the orbit is not expected to circularize before mass transfer starts because the circularization timescale for dynamical tides (relevant for a radiative envelope) in this case is extremely long 

\begin{align}
\tau_{\mathrm{circ}} &\sim \frac{1}{10} \left ( \frac{R_{Aa}^3}{G M_{Aa}} \right )^{1/2} \frac{1}{q(1+q)^{11/6}} E_2^{-1} \left ( \frac{a}{R_{Aa}} \right )^{21/2} \\ &\sim 370 \text{ Gyr} \left ( \frac{E_2}{10^{-7}} \right )^{-1} 
\end{align}

\noindent where $E_2$ is the second-order tidal coefficient that couples the tidal potential to gravity modes \citep{Zahn1977}. Even a significant enhancement of the dynamical tide through resonance between the tidal potential and gravity modes cannot make up for such a long circularization timescale. Therefore mass transfer should start at periastron before any significant tidal dissipation reduces the orbital eccentricity.  

In other words, the system is bound to go through ``eccentric Roche lobe overflow''. In contrast to circular binaries in which there is a rotating frame in which the equipotential surfaces are time independent, in an eccentric binary there is no such frame and the behavior of the mass transfer process for a significantly extended star is much more complex. Attempts to model the process using smoothed particle hydrodynamics simulations are still quite limited \citep[e.g.][]{Regos2005,Lajoie2011} and generally lead to significant mass loss through the outer Lagrangian points. For high eccentricity \cite{Regos2005} finds ``catastrophic'' behavior in which the donor quickly loses a large amount of its mass into a common envelope. In addition, there are additional effects not included in such simulations such as misaligned stellar spin axes, rapid rotation and resonant excitation of dynamical tides. Predicting the outcome of the eccentric mass transfer process in \textit{Nunki} is beyond the scope of this paper but it is fair to say that one of the possibilities is a merger into a $M \gtrsim 10 M_{\odot}$ star. Such a merger product would then be massive enough to eventually develop an iron core and undergo core collapse. 

There are two possible pathways to avoid a merger. The first one, as alluded above, is if the system could somehow circularize before becoming dynamically unstable. The second one is if the system were able to eject the common envelope before merging. The total orbital energy available before the core of the primary would merge with the secondary is 

\begin{align}
|\Delta E_{\mathrm{orb}}| &= \frac{G M_{Ab} M_{Aa,core}}{2 a_f} - \frac{G M_{Ab} M_{Aa}}{2 a_i} \\ &\simeq 2.4 \times 10^{48} \text{ erg}
\end{align}

\noindent where $M_{Aa,core} \simeq 0.9 M_{\odot}$ is the mass of the helium core of the primary, $a_f \sim R_{Ab} \simeq 7 R_{\odot}$ is the final separation and $a_i \simeq 1.26 \text{ au}$ is the initial separation. 

Meanwhile, the binding energy of the primary envelope is roughly 

\begin{align}
|E_{\mathrm{bind,env}}| &\simeq \frac{G M_{Aa} (M_{Aa}-M_{Aa,core})}{\lambda R_{Aa}} \\ &\simeq 9.0 \times 10^{48} \text{ erg} \left ( \frac{\lambda}{0.3} \right )^{-1} 
\end{align}

\noindent where $\lambda$ is a parameter that takes into account the structure of the envelope. From Table 1 in \cite{Dewis2000}, for a $M=6 M_{\odot}$ star expanded to $R \sim 50 R_{\odot}$, $\lambda = \lambda_g \sim 0.15$ when only including the gravitational binding energy and $\lambda = \lambda_b \sim 0.3$ when including all the thermal energy of the envelope as well. These values are lower than the typical $\lambda$ for a giant star because the donor is still relatively compact when the mass transfer process starts. 

Therefore, we conclude that $|E_{\mathrm{bind,env}}|$ is at least a few times higher than $|\Delta E_{\mathrm{orb}}|$. Even though there are additional energy sources other than orbital energy that could possibly be tapped into ejecting the envelope \citep[][and references therein]{Ivanova2013}, this favors a merger rather than an ejection outcome in case a common envelope is formed during catastrophic eccentric Roche lobe overflow.

\section{Conclusion} \label{sec:conclusion}

In this paper we have performed a deeper investigation into the bright nearby binary system \textit{Nunki} = Sigma Sagittarii. Our results can be summarized as follows:

\begin{enumerate}

\item By combining five interferometric measurements that fortunately provide a very decent orbital coverage, we determined the orbital parameters of the binary.

\item The resulting dynamical mass is consistent with the previously inferred isochrone masses within $2\sigma$. The Hipparcos parallax might be slightly biased by the photocenter motion of 0.4 mas, in which case the true distance is slightly closer at about 67 pc. 

\item Archival spectra do not show any evidence for binarity, confirming that equal mass binaries for which the radial velocity separation is many times lower than the intrinsic line widths are virtually undetectable spectroscopically. 

\item The low orbital inclination $i\simeq20^{\circ}$, the high $v \sin i \simeq 160 \text{ km}\text{ s}^{-1}$ and the lack of a decretion disk provide a strong hint for significant spin-orbit misalignment in the system. 

\item The significant eccentricity $e\simeq0.5$ is key for its future evolution as the system is bound to undergo the still very poorly understood process of eccentric Roche lobe overflow. 

\item We argue that this process could possibly lead to a common envelope event, which would most likely lead to a merger into a $M \gtrsim 10 M_{\odot}$ given that the envelope binding energy is at least a few times larger than the available orbital energy to eject it. 

\item \textit{Nunki} can therefore be considered the closest core collapse progenitor candidate, interestingly as the result of a merger event. 

\item We have also shown that \textit{Bellatrix}=Gamma Orionis, another core collapse progenitor candidate slightly farther away, is not a close equal mass binary as previously suspected. 

\end{enumerate}

\section*{Acknowledgments}

This research has made use of the CDS Astronomical Databases SIMBAD and VIZIER, NASA's Astrophysics Data System Bibliographic Services, NumPy \citep{van2011numpy} and matplotlib, a Python library for publication quality graphics \citep{Hunter2007}.

\section*{Data availability}
The data underlying this article is publicly available from the ESO archive.

\bibliographystyle{aasjournal}
\bibliography{main}{}

\appendix

\section{A. VLTI/GRAVITY observations of \textit{Bellatrix} ($\gamma$ Orionis)}
\label{app:Bellatrix}

$\gamma$ Ori was observed with VLTI/GRAVITY \citep{GRAVITY17} on 2024-09-12 (MJD=60565.38) using the four 1.8-m Auxiliary Telescopes (ATs) in the configuration A0-B5-J2-J6, with a largest projected baseline of 167.1 meters, corresponding to an angular resolution of 2.7 mas in the near-infrared (NIR) K band. The observations were performed in single-field mode, with half of the light used to track the fringes at high temporal resolution (0.85 ms) and low spectral resolution (R=22) and the other half integrated over 10s exposures at high spectral resolution (R=4,000). Two files were obtained, each containing 16 exposures. The seeing during the observations was 0.6" at 500 nm and the K2III star HD 34137 (angular diameter 0.74 mas) was observed as interferometric calibrator. 

\begin{figure*}[]
\centering
\includegraphics[width=\textwidth]{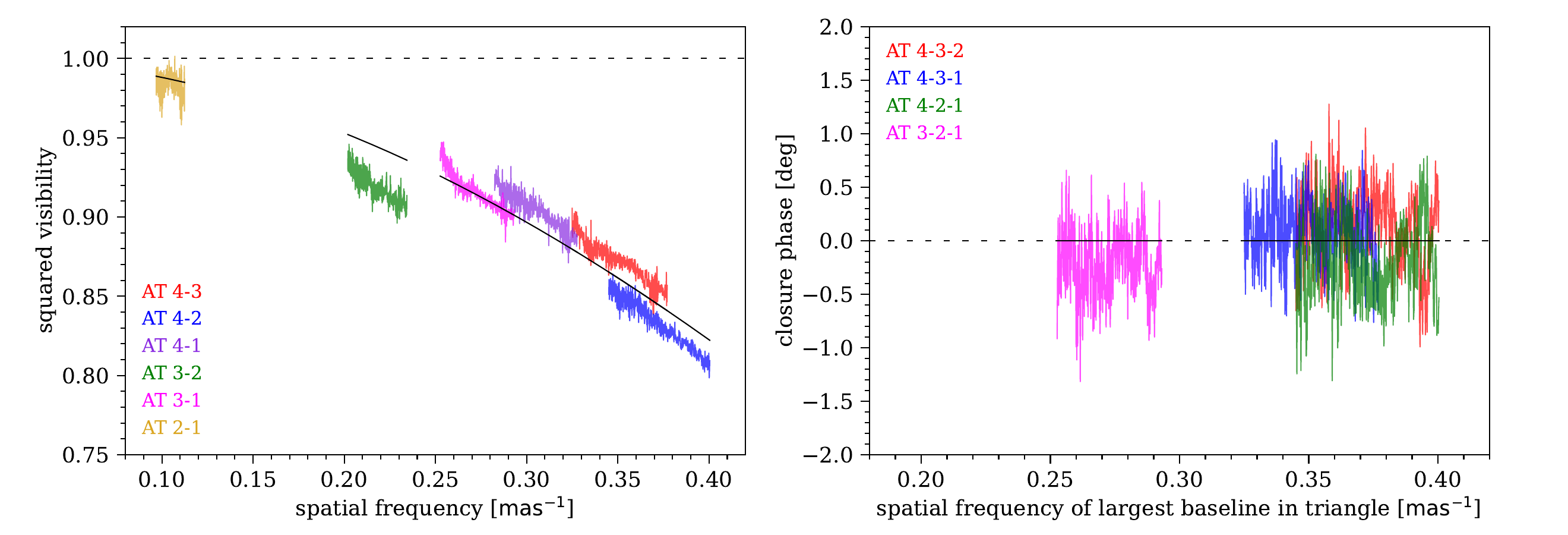}
\caption{\label{fig:gravity_bellatrix} VLTI/GRAVITY data (colored) for \textit{Bellatrix} = $\gamma$ Orionis. The black solid lines show the the best fit for a uniform disk that is partially resolved. The dashed lines show the expected values for a single unresolved star.}
\end{figure*}

\begin{figure*}[]
\centering
\includegraphics[width=\textwidth]{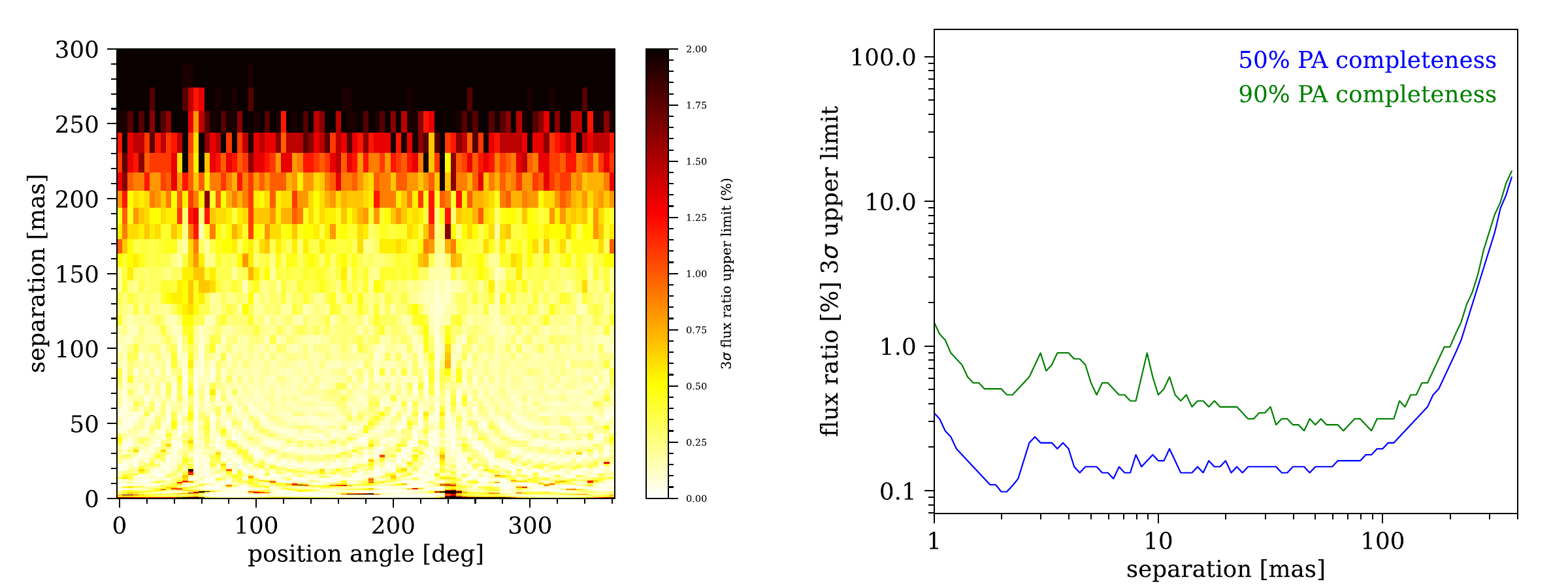}
\caption{\label{fig:bellatrix_flux_limit} K band flux ratio upper limit for a companion to \textit{Bellatrix} based on the closure phases of the VLTI/GRAVITY observations. The left panel shows the full 2D parameter space while the right panel shows the same data with 50\% and 90\% completeness in position angle.}
\end{figure*}

The interferometric data for one of the two files is shown in Figure \ref{fig:gravity_bellatrix}. The closure phases are consistent with zero within their RMS of $0.5^{\degr}$. Figure \ref{fig:bellatrix_flux_limit} shows the corresponding 3$\sigma$ K band flux ratio upper limit as a function of projected separation and position angle (left) and collapsed over position angle with 50\% and 90\% completeness (right). We can establish an upper limit on the flux ratio of about 1\% (corresponding to an absolute K band magnitude of 2.9 and a $1.1 M_{\odot}$ main-sequence star) over projected separations $1 \lesssim \rho \lesssim 200 \text{ mas}$. We conclude that \textit{Bellatrix} is \textit{not} a close equal-mass binary as was hinted in \cite{Nieva2012} based on its spectrum. 

The squared visibilities show that \textit{Bellatrix} is partially resolved with a best-fit angular diameter $\theta=0.72\pm0.02 \text{ mas}$ (solid black line in Figure \ref{fig:gravity_bellatrix} (left)). The small offsets between the data and the model are not due to intrinsic structure but rather due to small systematic errors in the squared visibility calibration. 

We made use of \texttt{MIST} isochrones \citep{Dotter16,Choi16,Paxton11,Paxton13,Paxton15} in order to estimate the mass of \textit{Bellatrix} based on its Tycho2 $V_T$ and $B_T$ magnitudes, its 2MASS K band magnitude and the Hipparcos distance and assuming solar metallicites and negligible extinction. For an initial rotation velocity of 0.4 times the critical velocity, we found a mass $M=8.4 M_{\odot}$, a temperature $T=20,000 \text{ K}$, a radius $R=6.2 R_{\odot}$ and an age of 28 Myr. The corresponding angular diameter $\theta=0.75 \text{ mas}$ is consistent with the measured one.

\section{B. VLTI/PIONIER data and binary fits for additional epochs}
\label{app:pionier_fits}

\begin{figure*}[]
\centering
\includegraphics[width=\textwidth]{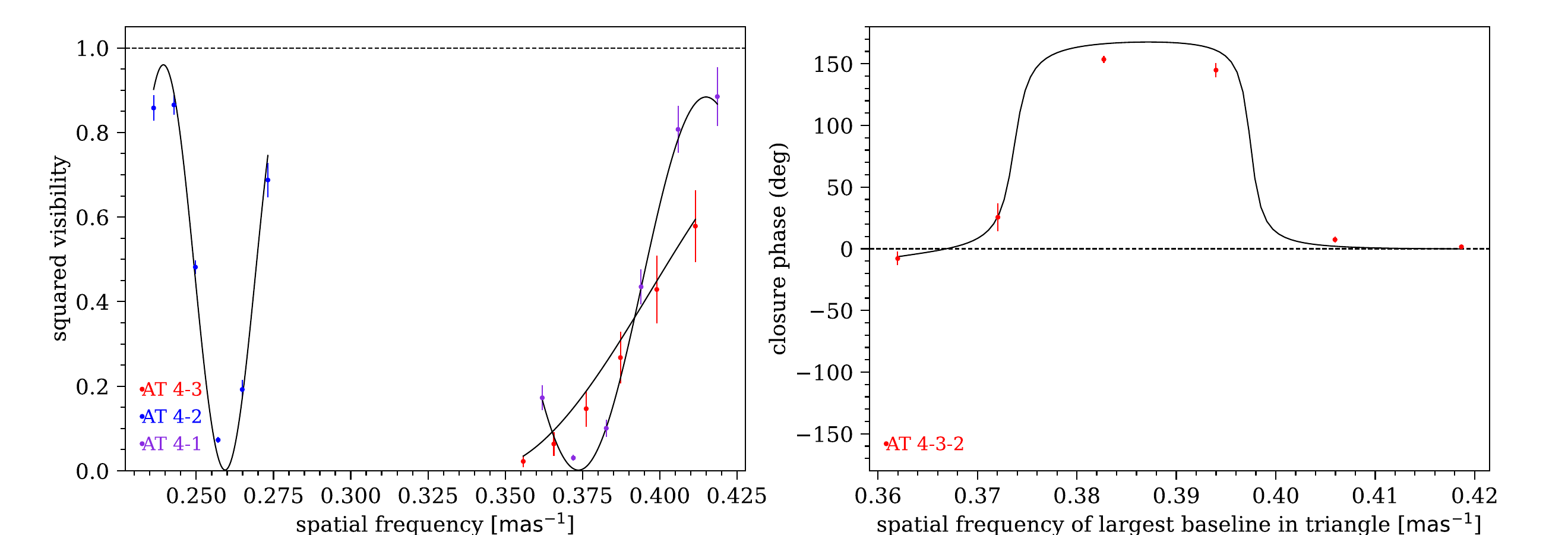} \\
\includegraphics[width=\textwidth]{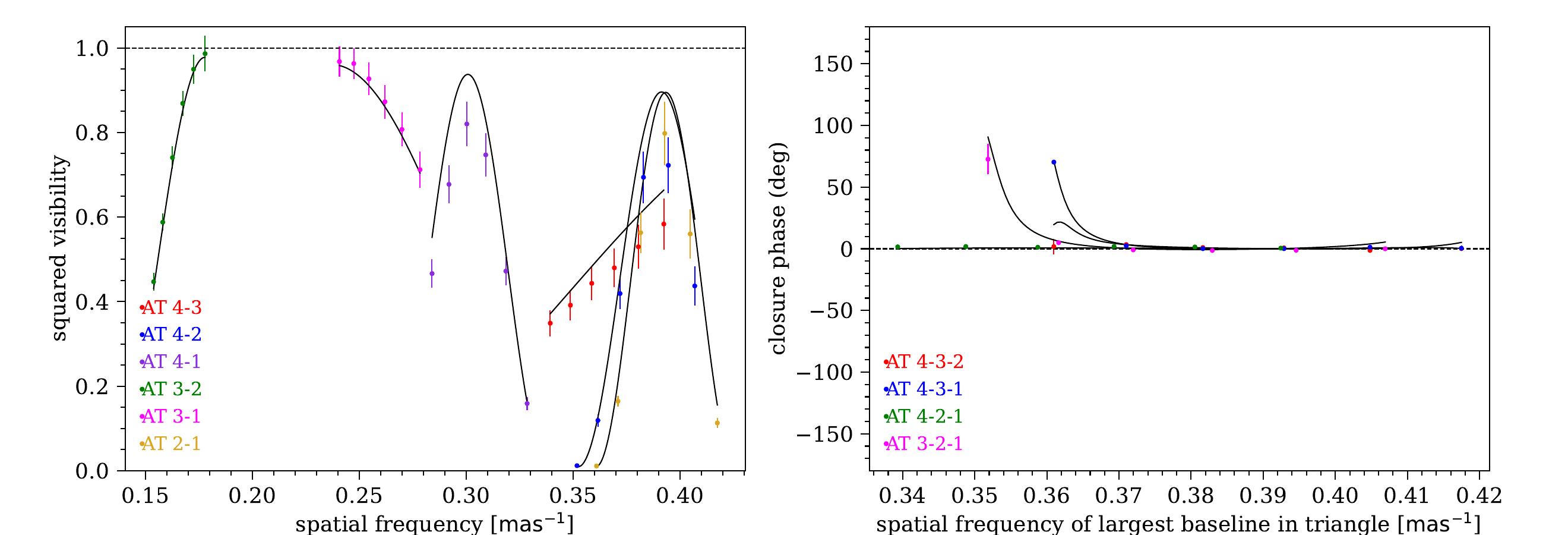}
\caption{\label{fig:pionier_fits_add} VLTI/PIONIER data (colored) and best fit binary model (solid black) for the \textit{Nunki} data in epochs 2017-07-16 (top) and 2017-09-09 (bottom). The dashed lines show the expected values for a single unresolved star.}
\end{figure*}

Figure \ref{fig:pionier_fits_add} shows the VLTI/PIONIER data (colored) and best fit binary model (solid black) for the epochs 2017-07-16 and 2017-09-09. 

\section{C. Parameter distributions for orbital fit}
\label{app:corner_plot}

\begin{figure*}[]
\centering
\includegraphics[width=\textwidth]{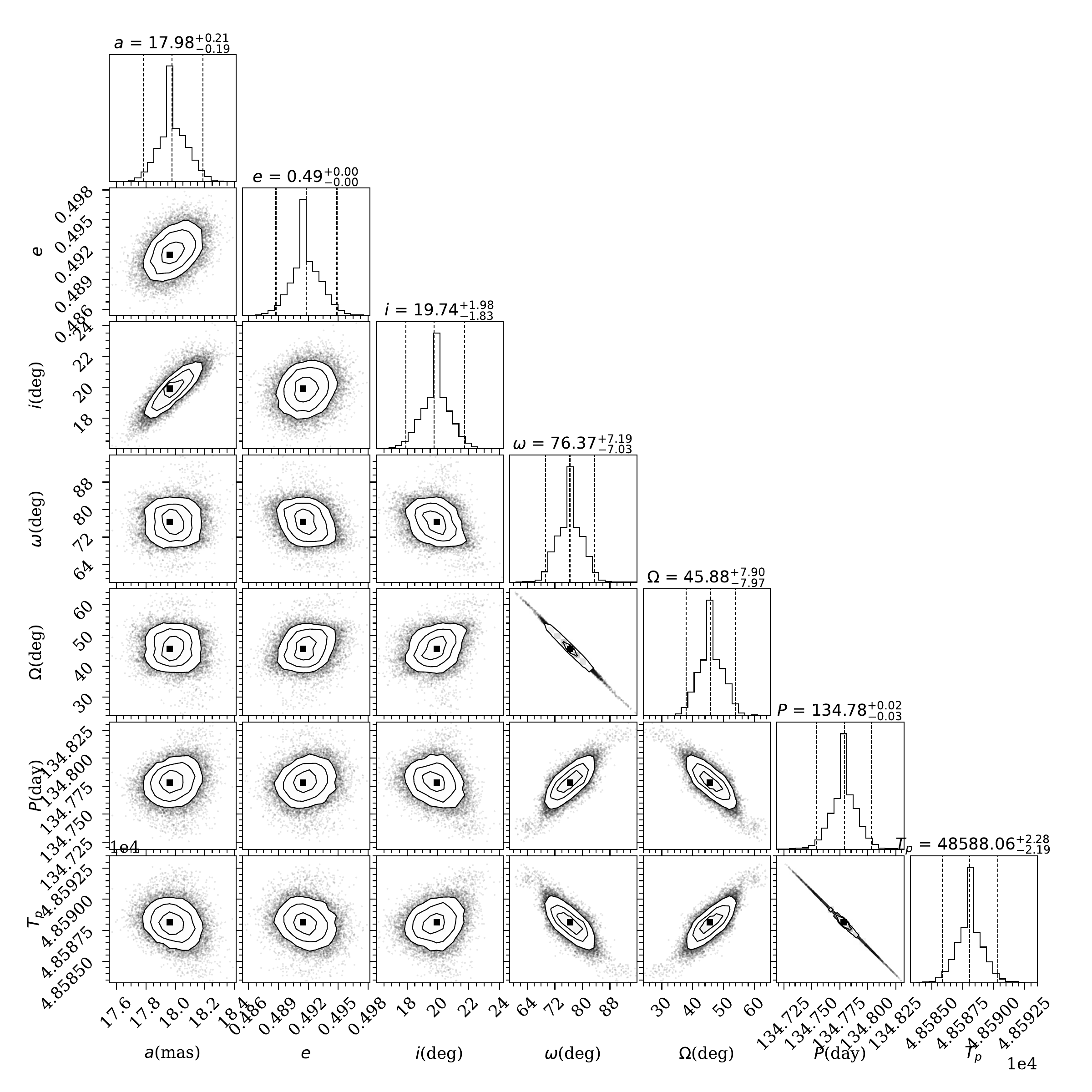}
\caption{\label{fig:corner} Orbital parameters distributions for \textit{Nunki}.}
\end{figure*}

The full distributions of best fit orbital parameters for \textit{Nunki} are shown in Figure \ref{fig:corner}. 

\section{C. Spectrum line list}

\clearpage

\newcommand{\Tstrut}{\rule{0pt}{2.8ex}}       
\newcommand{\Bstrut}{\rule[-1.2ex]{0pt}{0pt}} 

\begin{longtable}{c r @{\hspace{10pt}} c r}
\caption{\label{table:line_list} Identified spectral lines in \textit{Nunki.}} \\

\toprule
Ion & Wavelength (\AA) & & Wavelength (\AA) \\
\midrule
\endfirsthead

\toprule
Ion & Wavelength (\AA) & & Wavelength (\AA) \\
\midrule
\endhead

\bottomrule
\endfoot

\midrule
\multicolumn{4}{l}{\textbf{HI}} \\[4pt]

H16 \Tstrut\Bstrut & 3703.85 & H15 & 3711.97 \\
H14 \Tstrut\Bstrut & 3721.94 & H13 & 3734.37 \\
H12 \Tstrut\Bstrut & 3750.15 & H11 & 3770.63 \\
H10 \Tstrut\Bstrut & 3797.90 & H9 & 3835.38 \\
H8  \Tstrut\Bstrut & 3889.05 & H$\epsilon$ & 3970.07 \\
H$\delta$ \Tstrut\Bstrut & 4101.735 & H$\gamma$ & 4340.463 \\
H$\beta$ \Tstrut\Bstrut & 4861.325 & H$\alpha$ & 6562.80 \\
Pa16 \Tstrut\Bstrut & 8502.54 & Pa15 & 8545.44 \\
Pa14 \Tstrut\Bstrut & 8598.46 & Pa13 & 8665.08 \\
Pa12 \Tstrut\Bstrut & 8750.54 & Pa11 & 8862.85 \\

\midrule
\multicolumn{4}{l}{\textbf{HeI}} \\[4pt]

HeI \Tstrut\Bstrut & 3705.00 & & 3805.74 \\
HeI \Tstrut\Bstrut & 3819.60 & & 3833.57 \\
HeI \Tstrut\Bstrut & 3867.48 & & 3871.69 \\
HeI \Tstrut\Bstrut & 3926.41 & & 3964.73 \\
HeI \Tstrut\Bstrut & 4009.26 & & 4026.20 \\
HeI \Tstrut\Bstrut & 4120.82 & & 4143.76 \\
HeI \Tstrut\Bstrut & 4168.97 & & 4387.93 \\
HeI \Tstrut\Bstrut & 4437.55 & & 4471.50 \\
HeI \Tstrut\Bstrut & 4713.17 & & 4921.93 \\
HeI \Tstrut\Bstrut & 5015.70 & & 5047.74 \\
HeI \Tstrut\Bstrut & 5875.64 & & 6678.15 \\
HeI \Tstrut\Bstrut & 7065.71 & & \\

\midrule
\multicolumn{4}{l}{\textbf{CII}} \\[4pt]

CII \Tstrut\Bstrut & 3876.4 & & 3918.98 \\
CII \Tstrut\Bstrut & 3920.69 & & 4075.85 \\
CII \Tstrut\Bstrut & 4267.26 & & 6578.01 \\
CII \Tstrut\Bstrut & 6582.88 & & \\

\midrule
\multicolumn{4}{l}{\textbf{NII}} \\[4pt]

NII \Tstrut\Bstrut & 3995.00 & & \\

\midrule
\multicolumn{4}{l}{\textbf{OI}} \\[4pt]

OI \Tstrut\Bstrut & 7771.94 & & 7774.17 \\
OI \Tstrut\Bstrut & 7775.39 & & 8446.80 \\

\midrule
\multicolumn{4}{l}{\textbf{NeI}} \\[4pt]

NeI \Tstrut\Bstrut & 6402.25 & & \\

\midrule
\multicolumn{4}{l}{\textbf{MgII}} \\[4pt]

MgII \Tstrut\Bstrut & 4481.21 & & \\

\midrule
\multicolumn{4}{l}{\textbf{SiII}} \\[4pt]

SiII \Tstrut\Bstrut & 4128.05 & & 4130.88 \\
SiII \Tstrut\Bstrut & 5428.92 & & 5432.89 \\
SiII \Tstrut\Bstrut & 6347.09 & & 6371.38 \\

\midrule
\multicolumn{4}{l}{\textbf{SiIII}} \\[4pt]

SiIII \Tstrut\Bstrut & 4552.62 & & 4567.82 \\
SiIII \Tstrut\Bstrut & 4574.76 & & \\

\midrule
\multicolumn{4}{l}{\textbf{CaII}} \\[4pt]

CaII \Tstrut\Bstrut & 3933.66 & & \\

\end{longtable}

\end{document}